\documentclass[12pt,preprint]{aastex}

\usepackage{epsfig}
\usepackage{graphicx,color}

\newcommand{\beq}{\begin{equation}}
\newcommand{\eeq}{\end{equation}}

\begin{document}

\title{Physical Conditions in the Reconnection Layer in Pulsar Magnetospheres.}

\shorttitle{Reconnection in pulsar magnetosphere}

\author{ Dmitri~A.~Uzdensky\altaffilmark{1}} 
\author{ Anatoly Spitkovsky\altaffilmark{2}} 

\shortauthors{Uzdensky \& Spitkovsky}

\altaffiltext{1}{Center for Integrated Plasma Studies, Physics Department, University of Colorado, UCB 390, Boulder, CO 80309-0390}
\email{uzdensky@colorado.edu}

\altaffiltext{2}{Department of Astrophysical Sciences, Princeton University, Princeton, NJ 08544}
\email{anatoly@astro.princeton.edu}


\begin{abstract} 
The magnetosphere of a rotating pulsar naturally develops a current sheet beyond the light cylinder (LC).  Magnetic reconnection in this current sheet inevitably dissipates a nontrivial fraction of the pulsar spin-down power within a few LC radii.  We develop a basic physical picture of reconnection in this environment and discuss its implications for the observed pulsed gamma-ray emission. We argue that reconnection proceeds in the plasmoid-dominated regime, via an hierarchical chain of multiple secondary islands/flux ropes. The inter-plasmoid reconnection layers are subject to strong synchrotron cooling, leading to significant plasma compression. Using the conditions of pressure balance across these  current layers, the balance between the heating by magnetic energy dissipation and synchrotron cooling, and Ampere's law, we obtain simple estimates for key parameters of the layers --- temperature, density, and layer thickness. In the comoving frame of the relativistic pulsar wind just outside of the equatorial current sheet, these basic parameters are uniquely determined by the strength of the reconnecting upstream magnetic field. For the case of the Crab pulsar, we find them to be of order 10~GeV, $10^{13}\, {\rm cm^{-3}}$, and 10~cm, respectively. After accounting for the bulk Doppler boosting due to the pulsar wind, the synchrotron and inverse-Compton emission from the reconnecting current sheet can explain the observed pulsed high-energy (GeV) and VHE ($\sim 100$ GeV) radiation, respectively. Also, we suggest that the rapid relative motions of the secondary plasmoids in the hierarchical chain may contribute to the production of the pulsar radio emission. 
\end{abstract}

\keywords{Gamma rays: stars --- Magnetic reconnection --- Radiation mechanisms: non-thermal --- Relativistic processes ---  pulsars: general --- pulsars: individual (Crab)}


\section{Introduction}
\label{sec-intro}

\begin{figure}[h]
\epsscale{0.7}
\plotone{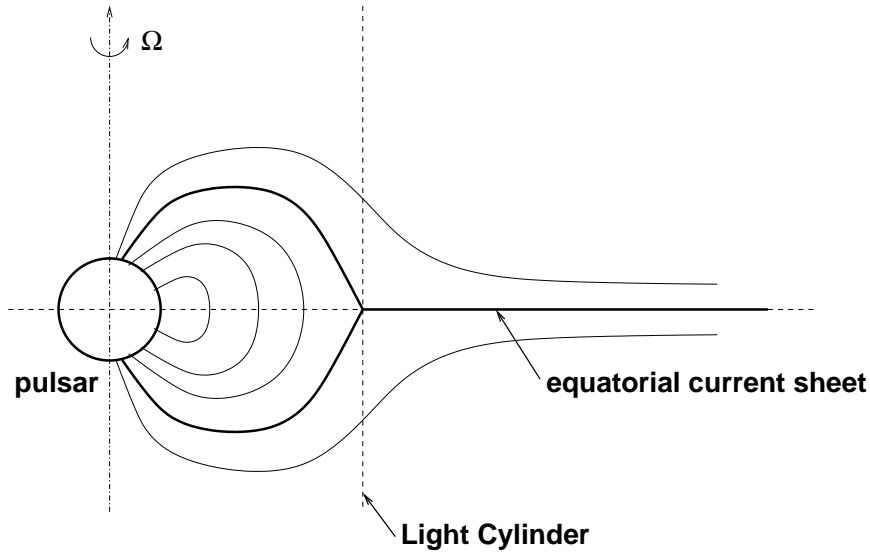}
   \caption{Basic structure of axisymmetric pulsar magnetosphere.
   \label{fig-1}}
\end{figure}

Simple theoretical reasoning shows that a plasma-filled magnetosphere of a rotating magnetic dipole, e.g., an axisymmetric pulsar, cannot remain closed beyond the rotator's Light Cylinder (LC) radius, $R_{\rm LC} = c \Omega^{-1}$, where $\Omega$ is the pulsar spin frequency \citep{Goldreich_Julian-1969}. 
Instead, the magnetic field lines extending beyond the LC have to open up, inevitably forming an equatorial current sheet (CS), see Fig.~\ref{fig-1}. The current sheet splits into two separatrix surfaces at the so-called Y-point, which is probably located at the~LC. This magnetospheric configuration has been studied extensively by a variety of numerical approaches, including relativistic force-free Grad-Shafranov equation \citep{Contopoulos_etal-1999, Gruzinov-2005, Timokhin-2006}, time-dependent force-free electrodynamic (FFE) simulations \citep{Komissarov-2006, McKinney-2006, Spitkovsky-2006,  Kalapotharakos_Contopoulos-2009, Kalapotharakos_etal-2012, Li_etal-2012, Petri-2012a, Parfrey_etal-2012},  and relativistic magnetohyrodynamic (MHD) simulations \citep{Komissarov-2006, Bucciantini_etal-2006}, and is now well understood. Moreover, its non-axisymmetric generalization, corresponding to an oblique pulsar, has also been obtained by time-dependent FFE simulations \citep{Spitkovsky-2006, Kalapotharakos_Contopoulos-2009, Kalapotharakos_etal-2012, Li_etal-2012, Petri-2012a}.

However, despite this remarkable progress in recent years, we cannot regard the above steady state  solutions as completely realistic. In particular, all of these numerical models suffer from one common problem --- they cannot properly treat the singular equatorial CS and have to resort to various artificial procedures to keep the codes stable. While the exact choices for these assumptions and procedures do not seem to have drastic effects on the overall gross structure of the magnetosphere, including such basic global parameters as the open poloidal magnetic flux fraction and the total spin-down power, they do affect strongly the behavior of the equatorial~CS, especially near the Y-point. We would like to stress that what happens in this region is important because it likely involves dissipation of a fraction of the outpouring electromagnetic energy via magnetic reconnection (see below). This is important for two reasons. First, reconnection alters magnetic field topology and thus affects the structure of the magnetosphere and of the pulsar wind \citep{Contopoulos-2005, Contopoulos_Spitkovsky-2006}; in particular, the bursty, non-stationary character of the reconnection process marked by plasmoid ejection is likely to lead to quasi-periodic motions in the magnetosphere 
\cite[see, e.g.,][]{Contopoulos-2005, Spitkovsky-2006, Bucciantini_etal-2006}. Secondly, reconnection in the equatorial CS (or ``ballerina's skirt" CS in the case of an oblique pulsar) means that a sizable fraction of the pulsar spin-down power is dissipated locally near the~LC, instead of contributing to the Poynting flux powering the pulsar wind at larger distances. A significant fraction of this dissipated energy is very likely to be radiated away promptly (see \S~\ref{sec-model}) and thus powers the observed high-energy gamma-ray emission \citep{Lyubarskii-1996, Lyubarskii-2000, Arons-2011, Arons-2012, Bai_Spitkovsky-2010, Contopoulos_Kalapotharakos-2010, Petri-2012b, Arka_Dubus-2012}. Figuring our the character (e.g., the spectrum) of this emission cannot be done completely within the framework of relativistic MHD or force-free models but requires a closer look at the basic plasma physics inside the reconnecting and radiating equatorial~CS. 

Determining these physical conditions is the main goal of the present paper. This includes both the description of the character of reconnection dynamics (magnetic structures and their motions), and the basic microscopic plasma parameters that determine the radiative cooling and the observable radiative signatures. 

\begin{figure}[h]
\epsscale{0.7}
\plotone{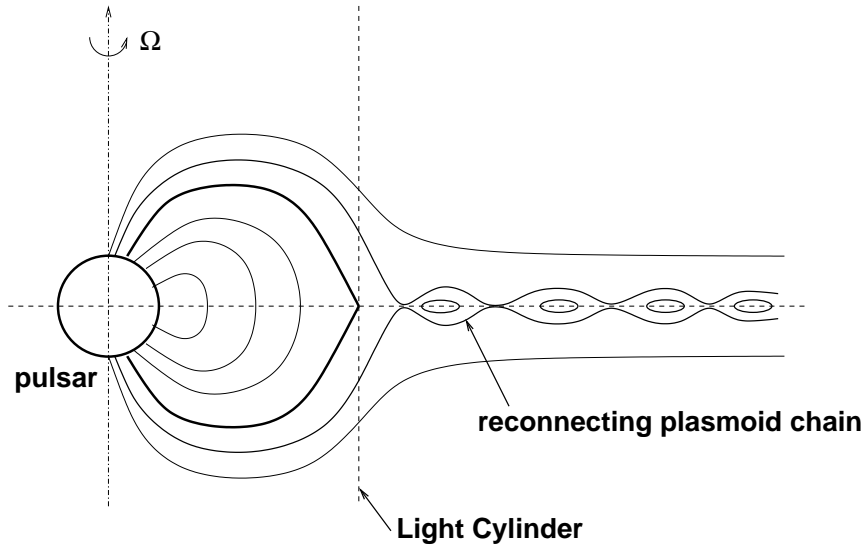}
   \caption{Tearing of the equatorial current sheet in pulsar magnetosphere
   \label{fig-2}}
\end{figure}

One issue that complicates the analysis of reconnection dynamics and, at the same time, provides a strong reason to believe that reconnection is unavoidable in these systems is the expected secondary tearing (aka plasmoid) instability of the equatorial current layer.  The resulting quasi-periodic reconnection and plasmoid ejection have been observed in numerical simulations \citep{Spitkovsky-2006, Bucciantini_etal-2006, Kalapotharakos_Contopoulos-2009}. 
Indeed, as the equatorial current layer forms just outside the LC and becomes thinner and thinner, at some point it will inevitably become unstable to tearing.  This happens when the aspect ratio of the current sheet --- the ratio of its length to its thickness --- exceeds a certain threshold which, as analytical arguments \citep{Bulanov_etal-1978, Loureiro_etal-2007, Bhattacharjee_etal-2009} and numerical simulations suggest, is probably of order~100 \citep{Samtaney_etal-2009, Daughton_etal-2009, Bhattacharjee_etal-2009, Huang_Bhattacharjee-2010, Loureiro_etal-2012, Komissarov_etal-2007, Liu_etal-2011, Zenitani_Hoshino-2008, Jaroschek_etal-2004a, Sironi_Spitkovsky-2011, Cerutti_etal-2012b}. As a result, the global current sheet breaks into a chain of rapidly growing secondary magnetic islands, separated by much smaller current sheets (see Fig.~\ref{fig-2}). Some of these current sheets may in turn themselves become so thin and long that they also undergo the same tearing instability, forming an hierarchy of smaller and smaller plasmoids and inter-plasmoid current sheets \citep{Shibata_Tanuma-2001, Bhattacharjee_etal-2009, Uzdensky_etal-2010}. This hierarchy truncates when the smallest elementary current sheets are no longer unstable to the plasmoid instability, i.e., again, when their aspect ratio is about~100.  
It is these elementary current sheets that are the actual sites of magnetic energy dissipation and its conversion to plasma energy (and, subsequently, radiation). Determining the basic physical plasma parameters in these elementary current sheets is critical for understanding how reconnection happens in pulsar magnetospheres and for interpreting its observable high-energy radiative signatures. This analysis is the main goal of the present paper. 

This paper is organized as follows. In Section~\ref{sec-model} we present the general estimates for the physical plasma parameters in the reconnecting CS in the pulsar magnetosphere. In Section~\ref{sec-implications} we apply this model to the Crab pulsar (\S~\ref{subsec-Crab}) and present the physical interpretation of the observed high-energy emission components in the framework of the reconnection model. We also discuss the conditions for the powerful pulsed synchrotron gamma-ray emission from the reconnecting CSs with implications for the observed gamma-ray turn-off line (\S~\ref{subsec-turnoff}). Finally, in Section~\ref{sec-conclusions} we present our conclusions.


\section{Basic plasma parameters inside a radiatively cooled reconnecting current layer in the pulsar magnetosphere}
\label{sec-model}

In this section we estimate the basic physical parameters of the plasma inside the smallest elementary inter-plasmoid current layers --- those at the bottom of the reconnection layer plasmoid hierarchy~\citep{Shibata_Tanuma-2001, Uzdensky_etal-2010}. The key distinguishing feature of our analysis from traditional reconnection theories is the inclusion of the strong optically thin synchrotron radiative cooling of the electrons and positrons inside the layer [see also \cite{Lyubarskii-1996, Petri-2012b}].   
An a priori reason why one may expect reconnection in pulsar magnetosphere (e.g., for the Crab) to be in the strong radiative cooling regime comes from noting that, for any reasonable plasma density and for a mega-gauss reconnecting magnetic field expected at these (light-cylinder) distances, the reconnection layer temperature, estimated from the pressure balance across the layer, is so high (in the GeV range), that the synchrotron cooling length is very small, measured in centimeters --- much smaller than the relevant system size. It is thus unavoidable that if reconnection does happen in this environment, it must be in the strong-cooling regime.

We start by outlining the key simplifying assumptions of our model. First, even though the plasma is collisionless with respect to Coulomb collisions and hence the particle distribution function may be very different from a (relativistic) Maxwellian, for simplicity in this paper we will ignore the fine details of the distribution function and will represent the average kinetic energy of the particles inside the layer by a single parameter that we will call the temperature $T= \gamma_T \, m_e c^2$, where $\gamma_T\gg 1$ is the thermal Lorentz factor. We are mostly interested in the perpendicular (to the upstream reconnecting magnetic field) temperature, since this is what enters both the pressure balance across the layer and the synchrotron radiative losses (see below). 

Next, we will assume that there is no guide magnetic field inside the reconnection layer.  In the case of reconnection via an equatorial current sheet of an axisymmetric pulsar considered here, this assumption is actually well justified by symmetry considerations --- i.e., we are dealing here with purely antiparallel reconnection. 

We are interested in reconnection occurring beyond the LC but not too far from it, namely, at distances comparable to the LC radius, $r-R_{\rm LC} \sim R_{\rm LC}$. In this region the reconnecting magnetic field upstream of the equatorial current layer has both radial and toroidal components which are comparable to each other. In the lab frame, there is also a poloidal electric field of comparable strength reversing across the equator.

The proper way to consider the reconnection problem is to do this in the frame of reference of the upstream plasma, i.e., the plasma just above and just below the current sheet \citep{Lyubarskii-1996}, which flies out along rotating open magnetic field lines with a relatively large bulk Lorentz factor of $\Gamma \sim 10-1000$ \citep{Lyubarsky-1995}.  Indeed, because in the lab (i.e., pulsar) frame the perpendicular (to the magnetic field) plasma flow is described by the ${\bf E\times B}$ drift, the upstream lab-frame poloidal electric field (which is of order the magnetic field and thus is not negligible in the vicinity of the~LC) vanishes in this comoving frame. In addition, the plasma upstream velocity is, of course, zero in the comoving frame by definition. These two simplifications leave us  with a clean canonical  reconnection problem setup where the plasma entering the reconnection layer is static (apart from the flows associated with the reconnection process itself).  In this approach, the upstream plasma is characterized only by three comoving-frame parameters: the upstream reconnecting magnetic field~$B_0$ (since there is no guide field), the density~$n_0$, and the pressure~$P_0$ (which we shall neglect in our analysis). Note that the comoving reconnecting field~$B_0$ should be comparable to the corresponding lab-frame magnetic field~\cite{Lyubarskii-1996}. This is because, although the bulk Lorentz factor of the upstream plasma flow may be large, $\Gamma \gg 1$, most of it is due to the ultra-relativistic parallel motion, which does not change the magnetic field strength, whereas the perpendicular plasma velocity, $|v_\perp| = c E_{\rm pol}/B_{\rm tot} = c\, (r/R_{\rm LC})\, B_{\rm pol}/B_{\rm tot}$, is only modestly relativistic at $r\sim R_{LC}$ (here $E_{\rm pol}$ and $B_{\rm pol}$ are the magnitudes of the poloidal components of the electric and magnetic field, respectively, and $B_{\rm tot}$ is the total magnetic field strength, all in the lab frame).  Another way to see that $B_0 \sim B_{0,\rm lab}$ is to use the fact that $B^2 - E^2$ is a Lorentz invariant and hence $B_0^2 = B_{0,\rm lab}^2 - E_{0,\rm lab}^2 \lesssim B_{0,\rm lab}^2$. 
On the other hand, we note that the comoving upstream plasma density~$n_0$ is much  lower (by a factor of~$\Gamma$) than the density in the lab frame.

Our main goal is to express the physical parameters in the layer --- the plasma density~$n$,  the temperature~$T$, and the thickness of the layer~$\delta$, all considered in the comoving frame, --- in terms of the upstream reconnecting magnetic field~$B_0$, and the reconnection rate, i.e., the reconnection electric field~$E$, which are treated here as given known quantities.  Our analysis is somewhat similar to that of~\cite{Lyubarskii-1996} and \cite{Petri-2012b}, although there are also many important differences (see~\S~\ref{sec-conclusions}). It also shares some similarities with the recent work by~\cite{Arka_Dubus-2012}.

The main equations governing the physical conditions inside the layer are the following. 
First, in the absence of a guide magnetic field (see above), the comoving frame  pressure balance across the layer (between the magnetic field outside and the plasma pressure inside, since we neglect the upstream plasma pressure) reads: 
\beq
2  n T = {{B_0^2}\over{8\pi}} \, , 
\label{eq-pressure-balance}
\eeq
where the factor 2 on the left-hand side accounts for the fact that we have a 2-species plasma (electrons and positrons): $n_e = n_{e^+} = n$. 

The second equation relating $T$, $n$, and $\delta$ is the Ampere's law in the direction of the main reconnection layer current (we ignore the displacement current although in principle it may lead to a modest correction):  
\beq
j = 2\, \beta_{\rm dr}\, e n c  \sim {c\over{4\pi}}\, {{B_0}\over{\delta}} \, ,  
\label{eq-Ampere}
\eeq
i.e.,
\beq
\delta \sim {B_0\over{8\pi\beta_{\rm dr}\, n e}} \, , 
\label{eq-Ampere-delta}
\eeq
where $c \beta_{\rm dr}$ is the average drift speed of current-carrying electrons and positrons inside the layer; we shall regard $\beta_{\rm dr}\lesssim 1$ as a constant parameter of order unity [in contrast with \cite{Arka_Dubus-2012} who assumed it to be small]. We shall also assume that $\beta_{\rm dr}$ is not too close to 1, i.e., regard $(1-\beta_{\rm dr})$ to be of order~1 as well.  Since the particles are ultra-relativistic, $\beta_{\rm dr}$ is the magnitude of the average cosine of the angle the particles make with the main current direction;  it is thus  a measure of angular anisotropy of the particle distribution function. We note that, as we shall show below, the assumption $\beta_{\rm dr} \sim 1$ is equivalent to the expectation that the layer thickness~$\delta$ be comparable to the typical Larmor radius of the particles in the layer, which is quite natural and common in collisionless reconnection.

Finally, the third key ingredient in our analysis is the energy conservation equation in the strong cooling regime.  This is described as a balance between the electro-magnetic energy (Poynting flux) flowing into the layer from both sides, $2\, S = 2\, (c/4\pi)\, E B_0$, and the radiative cooling rate per unit area, \citep[see, e.g.,][]{Lyubarskii-1996}.%
\footnote{Strictly speaking, this is not quite correct since, even in the strong radiative cooling regime considered here, a finite fraction of the energy may be carried away by the bulk kinetic and thermal energy of the reconnection outflow \citep{Uzdensky_McKinney-2011}; eventually, upon leaving the layer, this energy will be converted to the thermal (by viscosity and/or shocks) and residual magnetic energy of the plasmoids; the former will eventually be radiated away but because this happens already outside the elementary current layer under consideration, this radiation will have no effect on the thermal structure of the layer of interest to us here.  Thus, the energy leaving the layer results in a modification of our heating/cooling balance by a factor of order unity; for simplicity, however, we shall ignore this complication in our present rough analysis.}
Assuming radiative cooling to be optically thin, the latter can be estimated in terms of the volumetric cooling rate $Q$ as $(2\delta)\, Q \sim (2\delta)\, 2n\, \Lambda(B,T,n)$, where $\Lambda(B,T,n)$ is the cooling function (the radiative power per particle). Thus, the overall energy balance reads: 
\beq
{c\over 4\pi} \, E B_0 \sim 2n\, \Lambda(B,\gamma_T)\, \delta \, . 
\label{eq-energy}
\eeq

In the case of synchrotron radiation cooling considered here (we neglect IC and curvature radiation as cooling mechanisms although they may be important in producing the very high energy emission), the cooling function is 
\beq
\Lambda(B,\gamma) = 2 \, \sigma_T c \, {B_\perp^2\over{8 \pi}} \, \gamma^2 \, ,
\eeq
where $B_\perp$ is the magnetic field component perpendicular to the particle's motion, $\sigma_T = (8\pi/3) r_e^2$ is the Thomson cross section and $ r_e \equiv e^2/m_e c^2 \simeq 2.8 \times 10^{-13}\, {\rm cm}$ is the classical electron radius. 
For a rough estimate, we will assume that most synchrotron radiation is radiated by electrons with $\gamma\simeq \gamma_T$ in a perpendicular magnetic field $B_\perp \simeq B_0$ [although the field may be much lower for the most energetic particles focussed deep into the layer \citep{Kirk-2004, Uzdensky_etal-2011, Cerutti_etal-2012a}]. Then, our  energy balance yields: 
\beq
\gamma_T^2 \sim {{\beta_{\rm rec}}\over{\sigma_T n \delta}} \, ,
\label{eq-energy-synch-gamma}
\eeq
where we introduced a dimensionless reconnection rate parameter
\beq
\beta_{\rm rec} \equiv {v_{\rm rec}\over c}  = {E\over{B_0}} \, , 
\label{eq-beta_rec-def}
\eeq
as the ratio of the reconnection inflow velocity, $v_{\rm rec}$, to the light speed. 
We shall treat $\beta_{\rm rec}$ as a known finite number of order unity. Existing PIC simulations of relativistic pair reconnection suggest typical values  $\beta_{\rm rec} \sim 0.1-0.2$ \cite[e.g.,][]{Zenitani_Hoshino-2007, Zenitani_Hoshino-2008, Liu_etal-2011, Cerutti_etal-2012b, Kagan_etal-2012}. 
Note, however, that almost all of these simulations did not include the radiation reaction force which, we argue, actually plays a dominant role in pulsar magnetosphere reconnection, both as a cooling mechanism and as a strong source of  effective radiative resistivity. The only study of relativistic pair reconnection that does include synchrotron radiation reaction self-consitently, by \cite{Jaroschek_Hoshino-2009}, does not report a measurement of the reconnection rate. We note, however, that strong plasma compression expected in the strong radiative cooling regime considered here may have an accelerating effect on reconnection \citep{Uzdensky_McKinney-2011}.

Equations (\ref{eq-pressure-balance}), (\ref{eq-Ampere-delta}), and (\ref{eq-energy-synch-gamma}), represent a system of 3 equations for 3 unknowns $n$, $\gamma_T$, and~$\delta$. Solving this system we get: 
\begin{eqnarray}
\gamma_T^2 &\sim & {{\beta_{\rm rec}}\over{\sigma_T n\delta}}  \sim  
\beta_{\rm rec} \beta_{\rm dr} \, {{8\pi e}\over{\sigma_T B_0}} 
= 3\, \beta_{\rm rec} \beta_{\rm dr} \,   {\rho_0\over{r_e}}  \, , \\
\label{eq-T}
n &\sim & {1\over{\sqrt{12\beta_{\rm rec} \beta_{\rm dr}}}}\, {{B_0^2}\over{8\pi m_e c^2}}\, 
\sqrt{r_e\over{\rho_0}} \, ,  \\
\label{eq-n}
\delta &\sim & {2\over\beta_{\rm dr}} \, \gamma_T \rho_0 \, .
\label{eq-delta}
\end{eqnarray}
where we introduced the nominal cyclotron radius: 
\beq
\rho_0 \equiv {{m_e c^2}\over{eB_0}} \, .
\label{eq-def-rho_0}
\eeq

Let us now discuss the physical meaning of these results. 

1) The temperature comes out to be essentially (apart from the factor~$\beta_{dr}^{1/2}$  of order unity) at the classical synchrotron radiation-reaction limit 
$\gamma_{\rm rad}$ associated with the magnetic field~$B_0$, at which the accelerating electric force is balanced by radiative losses: $eE = eB_0 \beta_{\rm rec} = \Lambda(B_0,\gamma_{\rm rad})/c = 2 \sigma_T (B_0^2/8\pi)\, \gamma_{\rm rad}^2$ \citep{Guilbert_etal-1983, deJager_etal-1996, Aharonian_etal-2002, Lyutikov-2010, Uzdensky_etal-2011}. 
Another interesting implication of this estimate is that the average projection of the synchrotron radiation reaction force per particle on the direction of the reconnection electric field, which can be thought of as the effective radiative resistive drag force, is about  $\beta_{dr}^2$ times the electric force per particle, $eE = e B_0 \beta_{\rm rec}$. This implies that, even though here we are dealing with a collisionless plasma, the effective resistivity due to synchrotron radiation reaction provides an important (of order $\beta_{\rm dr}^2 \sim 1$) contribution in balancing the reconnection electric field in the generalized Ohm's law. 

2) As expected in reconnection, apart from factors of order unity, the current layer thickness $\delta$ is basically the relativistic Larmor radius corresponding to this temperature~(\ref{eq-T}), $\delta \sim \rho(\gamma_T) =\gamma_T \rho_0$; this, in turn, equals the total (electrons plus positrons) relativistic collisionless skin-depth $d = c/\omega_p = (\gamma_T\, m_e c^2/ 8\pi n e^2)^{1/2}$ by virtue of the pressure balance condition~(\ref{eq-pressure-balance}).

3) The layer's comoving plasma density $n$ is a function of $B_0$ only, independent of the upstream density. This is not surprising: in the strong-cooling regime with a small upstream plasma-$\beta$, the upstream ambient plasma entering the layer has to compress by a large factor in order to maintain the pressure balance with the upstream magnetic field \cite{Dorman_Kulsrud-1995, Uzdensky_McKinney-2011}. In particular,  the density inside the layer may be significantly higher than the expected upstream comoving density~$n_0$, in which case one expects an enhancement of the reconnection rate~\citep{Uzdensky_McKinney-2011}. Furthermore, as was suggested by \cite{Lyubarskii-1996}, the upstream plasma density may be increased by the pair production due to annihilation of high-energy $\gamma$-photons emitted by the layer. These secondary upstream pairs should be much cooler than the plasma in the reconnection layer and could provide a significant source of optical/UV/X-ray photons through their synchrotron emission~\citep{Lyubarskii-1996}.

It is interesting to note that all three quantities ($\gamma_T$, $n$, and~$\delta$) essentially depend only on the strength of the reconnecting magnetic field~$B_0$.  
It is convenient to normalize this magnetic field to a fundamental field scale, namely the classical critical field defined as a field at which the nominal $\rho_0$ becomes equal to the classical electron radius: 
\beq
B_{\rm cl} \equiv {{m_e c^2}\over{e r_e}} = {{(m_e c^2)^2}\over{e^3}} = {e\over{r_e^2}} \simeq 6.0 \times 10^{15} \, {\rm G} \, .
\eeq
[For reference, this field is $\alpha_{\rm fs}^{-1}  \simeq 137$ times larger than the critical quantum magnetic field $B_{\rm QED} \simeq 4.4 \times 10^{13}\, {\rm G}$, where $\alpha_{\rm fs}= e^2/\hbar c  \simeq 1/137$ is the fine structure constant.] Then, introducing a dimensionless field strength
\beq
b \equiv {B_0\over{B_{\rm cl}}} = {r_e\over{\rho_0}} \, , 
\eeq
we can rewrite the above expressions as follows (dropping numerical factors of order unity): 
\begin{eqnarray}
\gamma_T &\sim &  (\beta_{\rm rec} \beta_{\rm dr})^{1/2}\,  b^{-1/2} \, , \\
\label{eq-T-b}
n &\sim & (\beta_{\rm rec}\beta_{\rm dr})^{-1/2}\, {{B_0^2}\over{8\pi m_e c^2}}\, b^{1/2}
= (\beta_{\rm rec} \beta_{\rm dr})^{-1/2}\, (8\pi r_e^3)^{-1}\, \, b^{5/2}\, , \\
\label{eq-n-b}
\delta &\sim &  \sqrt{{\beta_{\rm rec}}/{\beta_{\rm dr}}}  \, r_e \,  b^{-3/2}\, .
\label{eq-delta-b}
\end{eqnarray}


\section{Observational Implications}
\label{sec-implications}


\subsection{Application to the Crab pulsar}
\label{subsec-Crab}

Let us now apply these results to the case of the Crab pulsar magnetosphere. 
We shall take the fiducial value of the magnetic field near the LC to be 
$B_{\rm Crab} = 4\, {\rm MG}$, so that $b_{\rm Crab} \simeq 6.7\times 10^{-10}$ and 
$\rho_0 = r_e/b \simeq 4.2\times 10^{-4}\, {\rm cm}$. We then obtain the following estimates for the key reconnection layer parameters (in the wind comoving frame): 
\begin{eqnarray}
\gamma_{T,\rm Crab} & \sim &   4 \times 10^4 \, (\beta_{\rm rec} \beta_{\rm dr})^{1/2}
\quad  \Rightarrow \quad  T = \gamma_T m_e c^2  \sim 
(\beta_{\rm rec} \beta_{\rm dr})^{1/2}\, 20\, {\rm GeV} \, ,  \\
\label{eq-gamma-Crab}
n_{\rm Crab} &\sim & (\beta_{\rm rec}\beta_{\rm dr})^{-1/2}\, 2\times 10^{13}\, {\rm cm^3}\, , \\
\label{eq-n-Crab}
\delta_{\rm Crab} &\sim & 2\, \sqrt{{\beta_{\rm rec}}/{\beta_{\rm dr}}}\, b^{-3/2} \, r_e 
\simeq 
1.2\times 10^{14} \, r_e \, \sqrt{\beta_{\rm rec}/\beta_{\rm dr}}   \simeq 
30 \, {\rm cm} \, \sqrt{\beta_{\rm rec}/\beta_{\rm dr}}  \, .
\label{eq-delta-Crab}
\end{eqnarray}

In the lab (pulsar) frame, the typical energy of the particles should then be in the range of hundreds of GeV and the density of order $10^{15}\, {\rm cm^{-3}}$, assuming the wind bulk Lorentz factor in this region of a few tens \citep{Lyubarsky-1995}.

Since the reconnection proceeds in the strong synchrotron radiation cooling regime, a significant fraction ($\sim1/2$) of the magnetic energy entering the reconnecting current sheet is promptly radiated away. The overall radiated power should basically scale with the total pulsar spin-down power~$L_{\rm sd}$, times the dimensionless reconnection rate parameter $\beta_{\rm rec}\sim 0.1$ times some geometrical factors of order unity; overall, it should be as high as a few percent of~$L_{\rm sd}$. 
Since the typical particles in the layer are essentially at the radiation reaction limit $\gamma_{\rm rad}$, the resulting synchrotron radiation comes out at $\epsilon_{\rm ph} \sim \beta_{\rm rec}\, 160\, {\rm MeV} \sim 10-20\, {\rm MeV}$ in the comoving frame, independent of the magnetic field \citep{deJager_etal-1996, Aharonian_etal-2002, Lyutikov-2010}; the emission is then sharply cut off at higher energies. Boosting this by a relativistic Doppler factor of a few tens, associated with the overall parallel flow,  results in an observed radiation in the GeV range. This is in general agreement with the characteristic break at a few GeV observed in the pulsed emission of the Crab and a few other pulsars \citep[e.g.,][]{Abdo_etal-2010a, Abdo_etal-2010b}. In addition, synchrotron emission in excess of the classical $\sim 160\, {\rm MeV}$ limit may be attributed to the efficient focusing of relativistic Speiser orbits of the most energetic particles deep into the reconnection layer and the resulting suppression of synchrotron radiation reaction \citep{Kirk-2004, Uzdensky_etal-2011, Cerutti_etal-2012a}.  In any case, we feel it is highly plausible that magnetic reconnection in the equatorial current sheet in the pulsar magnetosphere (at $r\sim R_{\rm LC}$) can be the main mechanism powering the observed pulsed gamma-ray emission, as was suggested by \cite{Lyubarskii-1996, Lyubarskii-2000, Bai_Spitkovsky-2010, Arons-2011, Arons-2012, Arka_Dubus-2012}, and also by \cite{Petri-2012b}, although the latter assumed that the radiation is produced in the striped wind at much larger distances.  

Furthermore, the typical {\it laboratory-frame} energies of the hot  ($T \sim 10\, {\rm GeV}$) electrons and positrons in the reconnection current layer are expected to be of order of hundreds of~GeV. Thus, these particles possess energy to produce the very-high-energy (VHE) pulsed emission recently discovered by~\cite{VERITAS-2011} and MAGIC \citep{Aleksic_etal-2011}.  This radiation is probably produced by inverse-Compton up-scattering of the UV or X-ray emission form the pulsar, similar to the model proposed by \cite{Bogovalov_Aharonian-2000, Aharonian_etal-2012}; however, in contrast to our model, in their picture the wind is cold. 
Alternative views were proposed by \cite{Lyutikov_etal-2012}, where the VHE radiation is attributed to synchrotron-self-Compton emission by the secondary plasma produced in the outer gap of the magnetosphere, and by \cite{Bednarek-2012} where it is produced by curvature radiation.

Calculating the actual observable light curves of the synchrotron HE emission and the IC VHE emission is beyond the scope of this paper. It requires  considering the problem in the actual three-dimensional geometry of the force-free magnetosphere of an oblique rotating pulsar and taking into account the special-relativistic light-delay effects, along the lines of the study by \cite{Bai_Spitkovsky-2010}.

Finally, it is interesting to think about the implications of the highly-dynamic, non-stationary plasmoid-dominated reconnection process for the {\it radio emission} of pulsars \citep{Arons-2012}, as well as for the overall structure of the pulsar wind.  First, note that $\delta\sim  \rho(\gamma_T) \sim 10\, {\rm cm}$  represents the smallest plasma scales in the comoving frame, at the bottom of the expected plasmoid hierarchy in the reconnection layer. The corresponding comoving length of the smallest current layers is probably about 100 times larger, i.e., on the order of 10~m. In the laboratory frame, these scales will be Lorentz-contracted to millimeters and tens of centimeters, respectively. The coherent plasma motions and structures on these scales (and larger, since we expect a self-similar hierarchy of plasmoids and current sheets) may contribute to powering the radio counterparts to $\gamma$-ray pulses (i.e., radio pulses that are in phase with the HE pulses), observed in some pulsars, such as the Crab \citep{Abdo_etal-2010b}. 
In addition, since here we are dealing with the case of reconnection without a guide field, the original current layer, as well as the secondary plasmoids (which in three dimensions look like cylindrical flux ropes), may become disrupted by the relativistic drift-kink instability (RDKI) \cite[e.g.,][]{Zenitani_Hoshino-2005}.  In principle, one may expect  RDKI to cause additional dissipation of the residual magnetic energy remaining in the secondary flux ropes and it also may introduce additional small-scale substructures contributing to observable radio emission. However, although the development and consequences of this instability in the present context of relativistic reconnection with strong radiative cooling are not yet completely understood, the only published study on this subject reports that tearing instability prevails over RDKI in the radiative regime \citep{Jaroschek_Hoshino-2009}.

As one moves out away from the pulsar, the reconnecting upstream magnetic field becomes mostly toroidal  and hence the secondary flux ropes and the main reconnection electric field become increasingly radially oriented. Furthermore, the secondary plasmoids/flux ropes gradually merge (coalesce) with each other, forming bigger structures. Eventually, however, in a radially expanding and accelerating relativistic wind, the neighboring ropes lose causal contact with each other and the coalescence process stalls, freezes out; then one gets a relatively small number of large, mostly radially elongated flux ropes. The resulting geometry is favorable for the mechanism for producing the observed bright X-ray knots on the so-called Chandra ring, proposed by~\cite{Arons-2012}.


\subsection{Implications for the Gamma-ray Turn-off Line}
\label{subsec-turnoff}

It is interesting to note that not all pulsars detected at radio frequencies also produce pulsed (GeV) emission detectable by FERMI. There seems to be a clear boundary in the $P-\dot{P}$ diagram below which radio pulsars do not shine in gamma rays. Following \citep{Petri-2012b}, we ask whether one can understand the reasons for this dichotomy in the context of the reconnection model. 

As described above, intense gamma-radiation at the characteristic GeV energies (tens of MeV in the wind comoving frame) are produced in the reconnection scenario by synchrotron radiation from the hot particles in the reconnection layer. This happens when the reconnection process is in the strong cooling regime, i.e., as long as the comoving synchrotron cooling time $\tau_{\rm sync}$ is shorter than the time  the particles spend in the reconnection region (in the wind comoving frame). Since we expect most of the energy dissipation to take place in the inner part of the magnetosphere, within a few light cylinder radii, we can take the characteristic length of the reconnection region to be of order $R_{\rm LC}$ in the laboratory frame. Then, in the wind comoving frame the characteristic time a particle spends in this region is about $t_{\rm travel}^{co} \sim  R_{\rm LC}/c\Gamma  =  P/2\pi \Gamma$. 
Thus, we can expect a strong HE emission (with a luminosity perhaps  as high as about 10\% of the overall spin-down power) if 
\beq
\tau_{\rm sync} <  P/2\pi \Gamma  \, .
\label{eq-condition}
\eeq

This condition can be recast in terms of the observable pulsar parameters ($P$, $\dot{P}$) as follows. 
The synchrotron cooling time, using our estimate~(\ref{eq-T}) for the plasma temperature, can be evaluated as
\beq
\tau_{\rm sync} \sim 
7\times 10^{-9}\, {\rm s} \, (\beta_{\rm rec} \beta_{\rm dr})^{-1/2}\, B_{\rm LC,6}^{-3/2}\, ,
\label{eq-tau_sync}
\eeq
where $B_{\rm LC,6}$ is the characteristic strength of the reconnecting magnetic field near the LC in mega-gauss.  Comparing with~(\ref{eq-delta}), we see that $\tau_{\rm sync}$ is by a factor $\beta_{\rm rec}^{-1} \sim 10$ greater than $\delta/c$. 

Since the magnetic field within the LC is approximately dipolar, $B_{\rm LC}$ can be estimated in terms of the equatorial magnetic field at the neutron star surface $B_{\rm NS}$ as $B_{\rm LC}\simeq B_{\rm NS} \, (R_{\rm NS}/R_{\rm LC})^3$, where  $R_{\rm NS} \simeq 10 \, {\rm km}$ is the neutron star radius. In turn, $B_{\rm NS}$ is related to the pulsar spin-down power $L_{\rm sd} = 4 \pi^2 I \dot{P} P^{-3}$, where $I \simeq 10^{45}\, {\rm g\, cm^2}$ is the pulsar moment of inertia, and hence to the pulsar spin deceleration rate $\dot{P}$ as 
$B_{\rm NS} \simeq 2.6 \times 10^{19}\, {\rm G} \, (P\dot{P})^{1/2}$. 
This results in the following estimate of the magnetic field at the~LC: 
\beq
B_{\rm LC} \simeq 3\times 10^8\, {\rm G}\, P^{-2.5} \, \dot{P}^{1/2} \, .
\label{eq-B_LC}
\eeq

Substituting this into our expression~(\ref{eq-tau_sync}) for $\tau_{\rm sync}$, we can write the condition (\ref{eq-condition}) for strong pulsed GeV gamma-ray emission from the reconnecting equatorial current sheet as 
\beq
 \dot{P}  \gtrsim  
 2\times 10^{-15} \, \Gamma^{4/3}\, (\beta_{\rm rec}\beta_{\rm dr})^{-2/3}\, P^{11/3} \sim 
10^{-12} \, \Gamma_2^{4/3}\, (\beta_{\rm rec}\beta_{\rm dr})^{-2/3}\, P^{11/3}  \, , 
\label{eq-turnoff-line}
\eeq
where $\Gamma_2$ is $\Gamma/100$, $P$ is given in seconds, and $\dot{P}$  is dimensionless. 

This relationship defines the so-called gamma-ray turn-off line in the $P-\dot{P}$ diagram. Most of the FERMI-detected pulsars, including millisecond pulsars, indeed lie on or above this line in the $P-\dot{P}$ diagram \citep{Abdo_etal-2010b}, in agreement with our expectation~(\ref{eq-turnoff-line}). 
The power-law slope in the above relationship (11/3) is close to what is observed; it is also close to the slope $d \log \dot{P}/d \log P = 4$ obtained by \cite{Petri-2012b} [see his equation (14)], which, however, appears to be based on somewhat different physical principles.



\section{Conclusions}
\label{sec-conclusions}

In this paper we put forward a basic physical picture of how magnetic reconnection happens in the equatorial current sheet just outside of the light cylinder in the pulsar magnetosphere. We argued that the current sheet is inevitably unstable to the tearing (aka plasmoid) instability (and perhaps also the relativistic drift-kink instability), which leads to a violent, highly dynamic magnetic reconnection process involving a hierarchical chain of secondary plasmoids/flux ropes of different sizes that are constantly forming and coalescing with each other. The actual magnetic energy dissipation takes place in many small elementary inter-plasmoid current sheets. Estimating the basic physical conditions within these reconnecting current sheets and discussing their observational implications constituted the main objectives of this paper. 

We argued that, in strong contrast to reconnection in more familiar solar-system environments, relativistically-hot reconnection layers in magnetospheres of pulsars such as the Crab are subject to strong radiative cooling. A large part of the magnetic energy flowing into the reconnection region as Poynting flux is promptly radiated away through synchrotron radiation.  The resulting heating/cooling balance condition, in combination with the condition of pressure balance across the reconnection layer (in the absence of a guide magnetic field) and Ampere's law, enabled us to estimate the key fundamental parameters of the layer --- the comoving pair plasma density~$n_0$, temperature~$T$, and the layer thickness~$\delta$ --- in terms of just a single dimensional system parameter, the (co-moving) reconnecting magnetic field~$B_0$, plus a couple dimensionless parameters of order unity, namely the dimensionless reconnection rate~$\beta_{\rm rec}$ and the average charge-carrier drift velocity in the layer, $\beta_{\rm dr}$. We then proceeded to obtain simple scalings of the plasma parameters with the dimensionless magnetic field~$b$ (i.e., $B_0$ normalized by the critical classical field~$B_{\rm cl} \simeq 6\times 10^{15}\, {\rm G}$).  

Our estimates show that the comoving plasma temperature inside the reconnection layer is essentially comparable to the classical synchrotron radiation-reaction limit (of order 10 GeV for the Crab), which implies that, in the wind comoving frame, most of the synchrotron radiation is emitted at energies of order $\beta_{\rm rec} \, m_e c^2 \, \alpha^{-1} \sim 160\, {\rm MeV} \, \beta_{\rm rec}$, where $\alpha \simeq 1/137$ is the fine structure constant. The comoving plasma density inside the layer required by the pressure balance is then essentially determined by the upstream magnetic field strength and can be much higher than, and relatively insensitive to, the ambient (upstream) density.  This density enhancement in the layer can be explained, for example, by the strong compression of the plasma entering the layer, expected in the strong cooling regime of reconnection. 
Finally, the characteristic comoving thickness $\delta$ of the current layer is expected to be comparable with the relativistic Larmor radius of electrons and positrons corresponding to the above temperature~$T$ (of order 30 cm in the comoving frame in the Crab case). Since the temperature is close to the radiation-reaction limit,  as we discussed above, this layer thickness is comparable, essentially by construction, to the corresponding synchrotron cooling length. 

These findings have important implications for our interpretation of the pulsed gamma-ray emission observed in many pulsars, including the Crab. First, as was also found by \cite{Lyubarskii-1996} and \cite{Petri-2012b}, the main synchrotron emission from the relativistically hot reconnection layer in the wind comoving frame is expected to be at around the classical synchrotron radiation-reaction limit of $\epsilon_{rad} \simeq 160\, {\rm MeV} \, \beta{\rm rec}$, where $\beta_{\rm rec} \sim 0.1-0.2$ is the dimensionless reconnection rate. After correcting for relativistic Doppler factor of order 100, associated with the overall ambient plasma flow, this synchrotron radiation is boosted into the GeV range and can thus account for the pulsed high-energy (GeV) emission detected by FERMI \cite{Lyubarskii-1996, Petri-2012b}. Furthermore, by imposing the strong-cooling condition (i.e., requiring that in the wind comoving frame the synchrotron cooling time of the hot electrons in the layer be shorter than their travel time across the inner part of the magnetosphere), we derived the condition [see eq.~(\ref{eq-turnoff-line})] for a given pulsar to be a strong HE emitter --- i.e., the $\gamma$-ray turn-off line in the $P\dot{P}$ diagram [see also \cite{Petri-2012b}].

Secondly, the inverse-Compton emission by the same electrons and positrons in the hot reconnection layer can produce VHE (hundreds of GeV) emission discovered recently in the Crab pulsar by \cite{VERITAS-2011} and MAGIC\citep{Aleksic_etal-2011}. Finally, the expected collective plasma motions of, e.g.,  plasmoids/flux ropes in the highly-dynamical reconnecting plasmoid chain seem to be in the range of scales at which they may contribute to the observed radio emission from the pulsar.

Our analysis in this paper has a lot of similarities with those presented by \cite{Lyubarskii-1996} and, more recently, \cite{Petri-2012b} and~\cite{Arka_Dubus-2012}.  However, there are also many important differences. In particular, we build our model upon the modern understanding of reconnection that incorporates the hierarchical picture of reconnection in the plasmoid dominated regime and the effects of strong radiative cooling and the associated strong plasma compression in the reconnection layer.  As a result, we are able to obtain explicit approximate expressions for the comoving plasma temperature, density, and layer thickness without invoking poorly known parameters such as the multiplicity in the upstream plasma, etc. 
Furthermore, whereas the studies by \cite{Lyubarskii-1996}, \cite{Petri-2012b}, and \cite{Arka_Dubus-2012} were mostly concerned with explaining the high-energy (GeV) pulsed emission, we also addressed the origin of the VHE (hundreds of~GeV) and radio emission and argued that reconnection in the pulsar magnetosphere may have important implications for both of them. 
Finally, whereas the \cite{Petri-2012b} model was applied for the pulsar wind at very large distances, $r\gg R_{\rm LC}$, we develop our model in the context of the inner part of the pulsar magnetosphere, the near-wind region $r\sim R_{\rm LC}$, as did \cite{Lyubarskii-1996} and \cite{Arka_Dubus-2012}.  The authors of the latter paper \citep{Arka_Dubus-2012}, however, are mainly interested in the case of a non-reconnecting current  sheet and do not discuss the reconnection rate and the corresponding reconnection-powered plasma heating. Also, they are mostly concerned with the case when radiative energy losses from the current layer are small, describing the strong-cooling case (which is the primary focus of our present study) as being beyond the scope of their work. Correspondingly, instead of using the energy balance between reconnection heating and radiative cooling to obtain a completely closed system of equations, they regard the ratio $\Delta$ of the current layer thickness  to~$R_{\rm LC}$ as an arbitrary free input parameter (which they typically take to be 0.01). In contrast, in our approach the thickness is determined self-consistently by the relevant reconnection-layer physics, including the heating/cooling balance.


\acknowledgements 
We thank Jonathan Arons, Andrei Beloborodov, Benoit Cerutti, Jeremy Goodman, Jason Li, and Yuri Lyubarsky for fruitful discussions. 
This work was supported by NSF grant PHY-0903851, DOE Grant DE-SC0008409, and NASA grants NNX10A039G and NNX12AD01G.


\bibliographystyle{apj}



\end{document}